\documentclass[aps,amssymb,amsmath, twocolumn, pra, superscriptaddress]{revtex4}
\usepackage{graphicx}
\usepackage{epsfig}
\usepackage{dcolumn}
\usepackage{bm}
\usepackage{bbm}
\usepackage[colorlinks, citecolor=red]{hyperref}
\usepackage{hyperref}
\usepackage{color}
\usepackage[up]{subfigure}
\newcommand{\be}{\begin{equation}}
\newcommand{\ee}{\end{equation}}
\newcommand{\bc}{\begin{center}}
\newcommand{\ec}{\end{center}}
\newcommand{\bea}{\begin{eqnarray}}
\newcommand{\eea}{\end{eqnarray}}
\newcommand{\ba}{\begin{array}}
\newcommand{\ea}{\end{array}}

\begin{document}
\title{Disorder induced localization and enhancement of entanglement in one- and two-dimensional quantum walks}
\author{C. M. Chandrashekar}
\email{c.madaiah@oist.jp}
\affiliation{Quantum Systems Unit, Okinawa Institute of Science and Technology, Okinawa 904-0495, Japan}

\begin{abstract}
The time evolution of one- and two-dimensional discrete-time quantum walk with increase in disorder is studied. We use spatial, temporal and spatio-temporal broken periodicity of the unitary evolution as disorder to mimic the effect of random medium. Disorder induces a dramatic change in the interference pattern leading to localization of the quantum walks. We show this analytically using the dispersion relation and group velocity of the DTQW dynamics in continuum limit.
In one-dimension, spatial disorder results only in a small decreases of the particle-position entanglement and counter intuitively, an enhancement in entanglement with temporal and spatio-temporal disorder is seen. This study signifies the Anderson localization of quantum state without compromising on the degree of entanglement could be implemented in a large variety of physical settings where quantum walks has been realized. The study presented here makes it feasible to explore, theoretically and experimentally the interplay between disorder and entanglement. This also brings up a variety of intriguing questions relating to its implications on algorithmic and other applications. 
\end{abstract}
\maketitle

\section{Introduction}
\label{intro}
Localization of all energy states of a single particle in disordered potential  is one of the remarkable predictions of quantum mechanics. This phenomenon originates from the broken periodicity in the dynamics of the systems due to disordered media leading the interference between multiple scattering paths to result in the absence of diffusion, localization\,\cite{LR85}. It was originally predicted in the context of transport of electors in disordered crystals by Anderson\,\cite{And58} and now has been experimentally observed and theoretically studied in a variety of systems\,\cite{KM93+}, including light waves\,\cite{AL85+}, matter waves\,\cite{DZS03+}, and Ising-type quantum network\,\cite{TJS02}. Quantum walks which are the quantum analog of classical random walk also evolve involving the interference of amplitudes of multiple traversing paths\,\cite{Ria58+, ADZ93}. Converse to interference leading to localization, interference in quantum walks results in a quadratically faster spread in position space when compared to its classical counterpart in one-dimension\,\cite{ABN01+}. Interference leading to faster diffusion of quantum walks have played a significant role in the development of quantum algorithms\,\cite{Amb03}. Subsequently,  it is widely used to simulate and explain dynamics in variety of the physical systems. For example, they have been used to explain the  mechanism of wavelike energy  transfer  within photosynthetic systems\,\cite{ECR07, MRL08}, to demonstrate the coherent quantum control over atoms\,\cite{CL08, Cha11a} and to explore topological phases\,\cite{KRB10}. Quantum walks have been categorized into two forms, continuous-time quantum walk (CTQW)\,\cite{FG98} and discrete-time quantum walk (DTQW) \,\cite{ADZ93, ABN01+}. Its experimental implementation in NMR system, ions, photons, and atoms\,\cite{DLX03+} have opened up a new dimension to simulate quantum dynamics in a physical systems like the recent demonstration of localization of photon's wavepacket\,\cite{SCP11}. Recently, studies on percolation of quantum walks\,\cite{LKB10, KKN12}  and quantum walks with connectivity alternation\,\cite{LPK11} using disorder processes have also been reported.  In an anticipation of continuation of  significant role of quantum walks in understanding and simulating dynamics in various naturally occurring and engineered systems in laboratory, its relevant to study the effect of disorder on the dynamics from different perspectives. 
\par
Disorder in general introduces randomness into the system. Some studies have shown the disorder in the DTQW evolution leading to deviation from quadratic speed-up to the localization of the interference of amplitudes of the multiple traversing paths around the origin\, \cite{RAS05, OKA05, JM10, YKE08, Kon10+, Cha11a}.    In particular, localization of the DTQW dynamics in one- dimension is shown by introducing drifts with constant momentum between two consecutive steps of the walk\,\cite{RAS05} or by evolving the walk in a random medium characterized by a static disorder\,\cite{YKE08}. The key factor for the interference effect to result in localization is the broken periodicity in the dynamics of the system induced by the disordered media. 
Broken periodicity, need not be mediated by a disordered or a random medium alone; operations  defining the dynamics of the system can also be made random to break the periodicity in the evolution such that they mimic the effect of a random medium in the system and manifest localization\,\cite{JM10, Cha11a}. Taking this into consideration, we introduce disorder into the one- and two-dimensional DTQW system using randomized unitary evolution and call it as a {\it randomized DTQW}. Unitary preserving randomized DTQW evolution operator is modeled in different forms: spatial disorder, temporal disorder and spatio-temporal disorder by assigning different quantum coin operation for each position, time, and both position-time, respectively.  From the dispersion relation and group velocity of the DTQW dynamics in continuum limit, we show that the disorder induced in the form of randomized unitary evolution affects the interference pattern leading to localization of the particle in position space. Our study is further supported by the numerical analysis of the dynamics by comparing the standard deviation of the probability distribution of the disordered evolution.  In addition, we also present the effect of disorder on the degree of entanglement between the particle and the position space. We show that the spatial disorder results in the small decreases of the particle-position entanglement and counter intuitively, an enhancement in entanglement with temporal and spatio-temporal disorder is seen. This signifies that the localization of quantum state without compromising on the degree of entanglement can be realized in a large variety of physical systems where DTQW has been experimentally implemented. However localization might have a negative effect on algorithmic applications which require faster spatial spread but can serve as a strong positive sign where localization of quantum states is required.  In addition, randomized unitary operations can assist localization of photons in photonic systems which can find applications in photon storage and quantum information processing.    
\par
In Sec.\,\ref{QWL}, we introduce the DTQW dynamics in the second order partial differential equation form using single and three parameter quantum coin operations and obtain the dispersion relations, phase velocity and group velocity for the wavepacket.  The dynamics with three different forms of disorder : spatial, temporal and spatio-temporal dependent quantum coin operations leading to localization of distribution is explained in Sec.\,\ref{disorder}. Localization in  two-dimension using two-state particle DTQW with disordered evolution is presented in Sec.\,\ref{2dLoc}. 
 Effect of disorder on the entanglement between the particle and position space during the DTQW evolution is shown in Sec.\,\ref{entDis} before concluding in  Sec.\,\ref{conc}. 

\section{Dispersion relation and group velocity of discrete-time quantum walk}
\label{QWL}
\subsection{Single parameter coin operation}

DTQW in one-dimension is modeled using the two-state particle on a position space.  The Hilbert space of the  particle,  ${\cal H}_c$ is defined by the internal states of the particle, $|\uparrow \rangle = \begin{bmatrix} 1  \\ 0 \end{bmatrix}$ and $|\downarrow \rangle =\begin{bmatrix} 0  \\ 1 \end{bmatrix}$ as the basis states and the position Hilbert space, ${\cal H}_p$ is defined by the basis states described in terms of $| x \rangle$, where $x \in {\mathbbm  I}$.  Each step of DTQW comprises of shift operator 
$S_x \equiv     \sum_x \left [  |\uparrow \rangle\langle
\uparrow|\otimes|x-1 \rangle\langle   x|   +  | \downarrow \rangle\langle
\downarrow|\otimes |x+1 \rangle\langle x| \right ]$ followed by the coin operation, 
\bea
\label{coinOP}
B (\theta)      \equiv  \begin{bmatrix} \begin{array}{clcr}
  \mbox{~~}\cos(\theta)      &     &     \sin(\theta)
  \\ -\sin(\theta) & &  \cos(\theta) 
\end{array} \end{bmatrix}
\eea
which evolves the particle into the superposition of its basis states. Therefore, the effective operation for each step of the DTQW to evolve the particle in superposition of the position space is 
\be
\label{Wop}
W_x (\theta)\equiv  [B (\theta) \otimes  {\mathbbm 1}] S_x.
\ee
The state after time $t$ ($t$ steps), $|\Psi_t\rangle=W (\theta)^t|\Psi_{\rm in}\rangle$,
where  
$|\Psi_{\rm in}\rangle= \left ( \cos(\delta/2)| \uparrow \rangle + e^{i\eta}
\sin(\delta/2)|\downarrow \rangle \right )\otimes | 0\rangle$,
is the initial state of the particle at a position $x=0$. The coin parameter $\theta$ controls the variance of the probability distribution in position space. The iterative form of the state of the particle at each position $x$ and time $t+1$ in the form of the left- moving ($\psi^{\uparrow}$) and right- moving ($\psi^{\downarrow}$) components are, 
\bea
\label{eq:3}
\begin{bmatrix}   \begin{array}{clr}
 \psi^{\uparrow}_{x, t +1}  \\
\psi^{\downarrow}_{x, t +1} 
\end{array} \end{bmatrix}
  =  \begin{bmatrix} \begin{array}{clcr}
  \mbox{~}\cos(\theta)      &     &     \sin(\theta)
  \\ -\sin(\theta) & &  \cos(\theta) 
\end{array} \end{bmatrix} \begin{bmatrix}   \begin{array}{clr}
 \psi^{\uparrow}_{x+1, t }  \\
\psi^{\downarrow}_{x-1, t } 
\end{array} \end{bmatrix}.
\eea
Decoupling the basis states, $\psi^{\uparrow}$ and $\psi^{\downarrow}$ we obtain a two identical expression which can be represented by a single expression of the form, 
\bea
 \label{eq:4}
\psi_{x, t +1} + \psi_{x, t -1} 
= \cos(\theta) \Big [ \psi_{x-1, t}  +  \psi_{x+1, t} \Big ].
\eea 
Subtracting both side of the Eq.\,(\ref{eq:4}) by $2\left [ 1 + \cos(\theta)\right ]\psi_{x, t}$, we obtain a difference form of the expression that can be written as a second order partial differential equation \,\cite{CSB10},
\bea
\label{eq:5}
\label{KGe}
\Bigg[  \frac{\partial^2}{\partial t^2} - \cos(\theta) \frac{\partial^2}{\partial x^2} + 2  [1 - \cos(\theta)] \Bigg ] \psi_{x, t} = 0.
\eea
For the preceding partial differential equation governing the dynamics of the DTQW in continuum limit, one can seek a Fourier-mode wave like solution of the form 
\be
\label{eq:sol}
\psi_{x, t} =  e^{i(k x - \omega t)},
\ee
where $\omega$ is the wave frequency and $k$ is the wave number.
\begin{figure}[ht]
\bc 
\subfigure[]{\includegraphics[width=9.5cm]{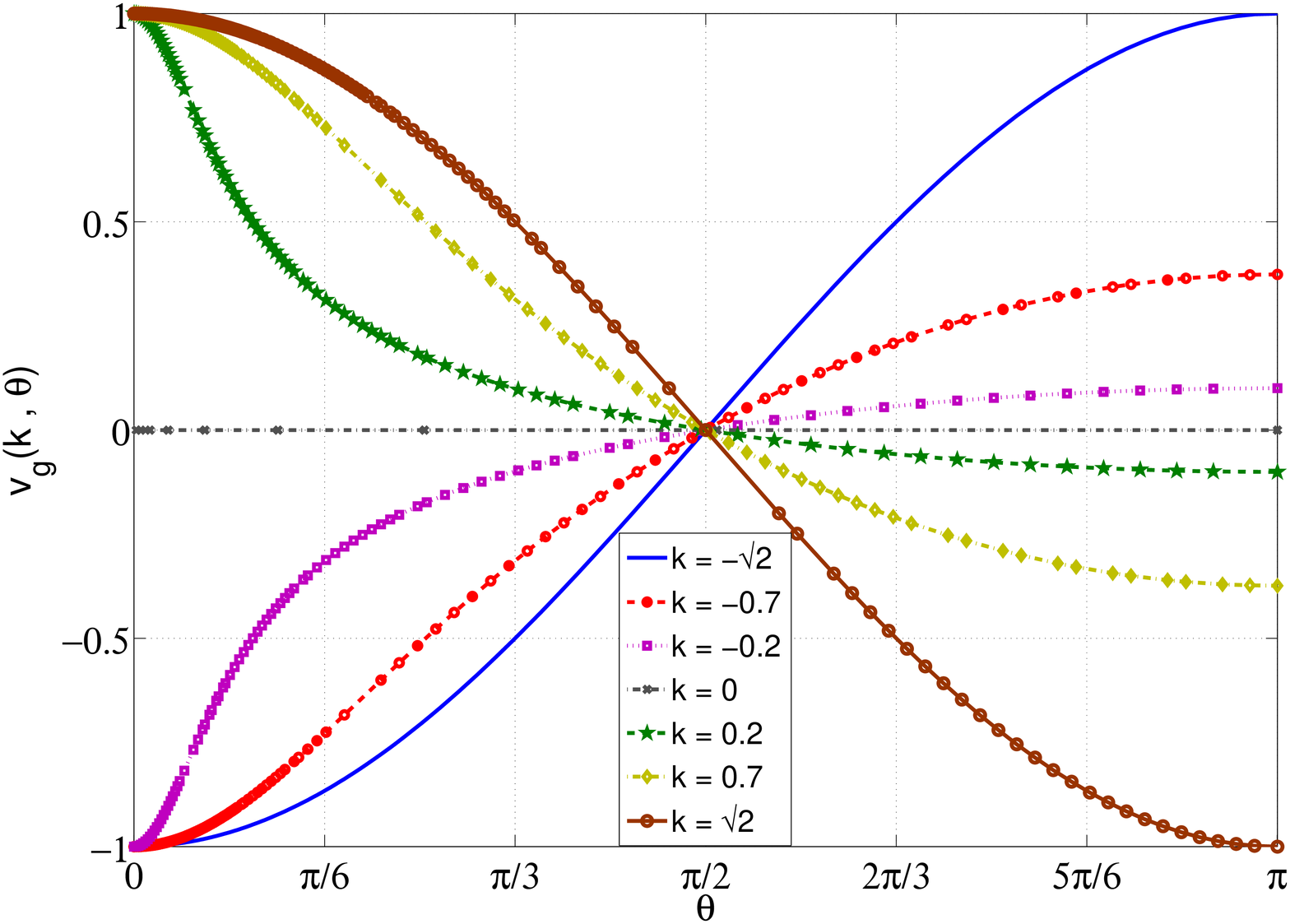} 
\label{fig:1a}}
\vskip -0.07in 
\subfigure[]{\includegraphics[width=9.5cm]{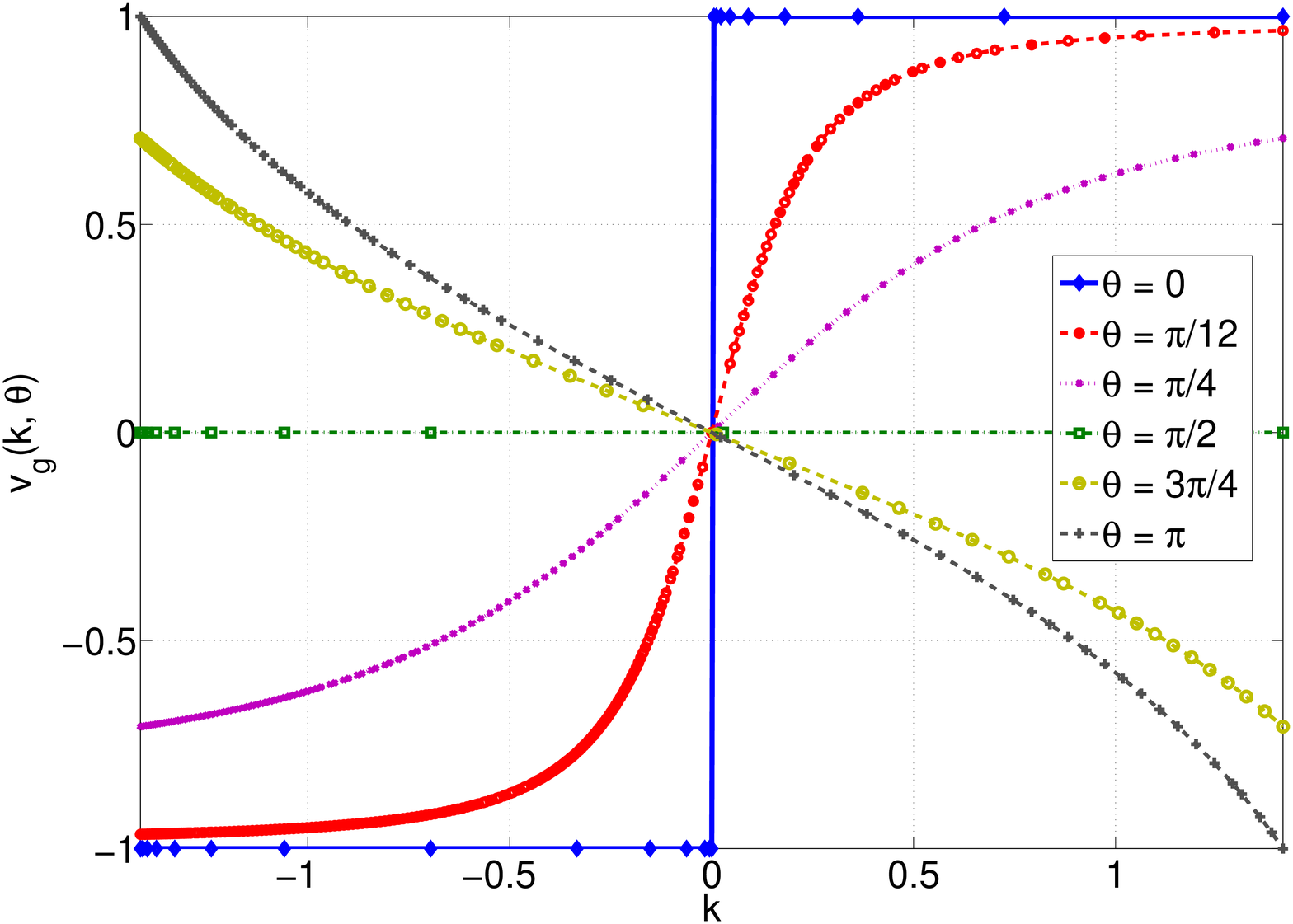} 
\label{fig:1b}}
\ec
\vskip -0.7cm
\caption{\footnotesize{(Color online) Group velocity as a function of  (a) $\theta$ for different values of $k$ and (b) $k$ for different value of $\theta$.  The $|v_g(k, \theta)|$ is maximum when $\theta=0$  for all value of $k$ and when $\theta = \pi$ only when $k = \pm \sqrt{2}$.  \label{fig:1}}}
\end{figure}
Upon substitution of the partial derivatives of Eq.\,(\ref{eq:sol}) into Eq.\,(\ref{eq:5}), the problem reduces to solving an algebraic equation,
\bea
\label{dispersion1}
\omega^2 -k^2 \cos(\theta) + 2[1 - \cos(\theta)]=0,
\eea
 which is called the dispersion relation. Solving for $\omega$ we obtain,
 \be
 \label{omega}
 \omega(k, \theta) = \pm \sqrt{k^2\cos(\theta) + 2[1-\cos(\theta)]},
 \ee
and  for the traveling wave to exist, the value of $k$ and $\theta$ should satisfy,
\be
\label{wavesol}
k^2\cos(\theta) + 2[1-\cos(\theta)]  > 0.
\ee
The {\it phase velocity} and {\it group velocity} for the solution of the form $e^{i(k x - \omega t)}$ has the natural definition $v_p := {\mathfrak R}[\omega]/k$  and $v_g := d{\mathfrak R}[\omega]/dk$, respectively. Therefore, the wavepackets $v_p$ and $v_g$ of each basis states, $\psi^{\uparrow}$ and $\psi^\downarrow$ during the DTQW evolution are,
 \bea
 \label{eq:vp}
 v_p (k, \theta) \equiv \frac{\omega(k, \theta) }{k}  = \pm \frac{\sqrt{k^2\cos(\theta) + 2[1-\cos(\theta)]}}{k}~,~
 \eea
 \bea
 \label{eq:vg}
 v_g (k, \theta) \equiv \frac{d \omega(k, \theta)}{d k} =  \pm \frac{k\cos(\theta)}{\sqrt{k^2 \cos(\theta) + 2 [1-\cos(\theta)]}}.
 \eea
In the coin operation $B(\theta)$ [Eq.\,(\ref{coinOP})], $\theta$ can be in the range : $0 \leq \theta \leq \pi$ and for any $\theta$ in the range : $\pi/2 < \theta \leq \pi$, $\cos(\theta)$ will be negative. This should be interpreted as the displacement of the left- moving component ($\psi^{\uparrow}$) to the right and the right- moving component ($\psi^{\downarrow}$) to the left. Therefore, for fixed $\theta$, the effective displacement (spread) of the wavepacket after time $t$ will be 
\be
D(t) = t |v_g(k, \theta)|.
\ee
\par
For both the extreme values, $\theta=0$ and $\theta=\pi$ the two states move away from each other without any interference, that is, $v_p = v_g$. By substituting $\theta =0$ in  Eqs.\,(\ref{eq:vp}) and (\ref{eq:vg}) we obtain $v_p = v_g = 1$ (maximum velocity) and for $\theta = \pi$, $v_p = v_g =1$ when we set $k=\sqrt{2}$.  Therefore, for evolution using the coin operation $B(\theta)$, the value of $k$ can be any value in the range: $-\sqrt {2} \le k \leq \sqrt {2}$.
\par
In Fig.\,\ref{fig:1}, we show the group velocity as a function of $\theta$ for different values of $k$ (Fig.\,\ref{fig:1a}) and as a function of $k$ for different value of $\theta$ (Fig.\,\ref{fig:1b}).  For negative value of $k$, an increase in  $v_g(k, \theta)$ with increase in $\theta$ is seen and for positive value of $k$, a decrease in  $v_g(k, \theta)$ with increase in $\theta$ is seen.  Absolute value of group velocity is maximum ($|v_g(k, \theta)| =1$) for all non-zero $k$ only when $\theta = 0$. An other value of $\theta$ where we can see $|v_g(k, \theta)| =1$  is for  $\theta = \pi$ only when $k = \pm \sqrt{2}$.  With increase in $\theta$ from $0$ to $\pi/2$ we see a decrease in $|v_g(k, \theta)|$ to zero and  further increase in $\theta$ from $\pi/2$ to $\pi$ results in an increase in $|v_g(k, \theta)|$. 
 
\subsection{SU(2) coin operation}
\label{su2gv}

The coin operation $B (\theta)$ [Eq.\,(\ref{coinOP})] in general can be written as an arbitrary SU(2) operator parameterized using three Euler angles,
\bea
\label{SU2coin}
B (\xi, \theta, \zeta)      \equiv  \begin{bmatrix} \begin{array}{clcr}
  \mbox{~~}e^{i(\xi + \zeta)}\cos(\theta)      &     &    e^{i(\xi - \zeta)} \sin(\theta)
  \\ -e^{-i(\xi -\zeta)}\sin(\theta) & &  e^{-i(\xi+\zeta)}\cos(\theta) 
\end{array} \end{bmatrix}.
\eea
Here the Euler angles $\xi$, $\theta$, $\zeta$  are restricted to the ranges:  $0 \leq \xi < 2\pi$ ; $0 \leq \theta \leq \pi$ ; $0 \leq \zeta < 2\pi$.  
Using $B(\xi, \theta, \zeta)$ in place of $B(\theta)$ we obtain the state of the particle at position $x$ and time $t+1$ in the form of $\psi^{\uparrow}$ and $\psi^{\downarrow}$. After decoupling basis states $\psi^{\uparrow}$ and $\psi^{\downarrow}$ we obtain two expressions identical to each other in the form, 
\bea
 \label{su2decop}
\psi_{x, t +1} + \psi_{x, t -1} =  \cos(\theta) \Big [e^{-i(\xi + \zeta)} \psi_{x-1, t}  + e^{i(\xi + \zeta)} \psi_{x+1, t} \Big ].~~
\eea
Subtracting both the side of the Eq.\,(\ref{su2decop}) by $[2\psi(x, t) + 2\cos(\theta) \cos(\xi + \zeta) ]\psi(x, t)$ we obtain a difference form of the expression that can be written as a second order partial differential equation,
\bea
\label{KGe1}
\Bigg[  \frac{\partial^2}{\partial t^2} - \cos(\theta) \Bigg \{ \cos(\xi + \zeta)\frac{\partial^2}{\partial x^2}   + 2 i  \sin(\xi + \zeta)\frac{\partial}{\partial x}\Bigg \}   \nonumber \\
+ 2 \Big [1 - \cos(\theta)\cos(\xi + \zeta) \Big ] \Bigg ] \psi_{x, t} = 0.
\eea
The preceding differential equation governing the dynamics in continuum limit models mixtures between particle wave propagation's dependency on $\theta$, $\xi$ and $\zeta$ and their effect on the interference patterns.  Upon substitution of the partial derivatives of the Fourier-mode wave like solution of the form given by Eq.\,(\ref{eq:sol})  into Eq.\,(\ref{KGe1}), we obtain the dispersion relation,
\bea
\omega^2  = k^2 \cos(\theta) \cos(\phi) + 2 k \cos(\theta)\sin(\phi)  \nonumber \\
- 2\Big[  1-\cos(\theta) \cos(\phi) \Big],
\eea 
where $\phi = \xi + \zeta$. The solution for $\omega$ as a function of $k$, $\theta$ and $\phi$ is,
\bea
\omega(k, \theta, \phi) =  \pm \sqrt{ \cos(\theta) \cos(\phi)[k^2 -2] + 2[1 -  k \cos(\theta)\sin(\phi)] }.~~~
\eea
For the values of $k$, $\theta$ and $\phi$ satisfying 
\be
\cos(\theta) \cos(\phi)[ k^2 -2] + 2[1 -  k \cos(\theta)\sin(\phi)] >0,
\ee
the phase and group velocity  are
\bea
\label{eq:17}
v_p(k, \theta, \phi) = \frac{\omega(k, \theta, \phi) }{k},
\eea
\bea
\label{eq:18}
v_g(k, \theta, \phi) = \frac{   \cos(\theta) [k\cos(\phi) - \sin(\phi)]}{ \omega(k, \theta, \phi) }.~~
\eea
Therefore, the displacement (spread) of the wavepacket for each time of the DTQW using arbitrary SU(2) coin operation will be $|v_g(k, \theta, \phi)|$  and the displacement after time $t$ will be, 
\be
D(t) = t |v_g(k, \theta, \phi)|.
\ee

\section{Disorder evolution and localization} 
\label{disorder}
 We introduce disorder into the DTQW evolution in different forms: using randomized quantum coin operation for each position (spatial disorder),  each time step (temporal disorder) and both position and time (spatio-temporal disorder). 
\subsection{Spatial disorder}
\label{td}

We define the spatial disorder in the DTQW evolution by introducing a position dependent coin operation $B(\theta_x)$  or $B (\xi_x, \theta_x, \zeta_x)$, where  
$\xi_x$, $\theta_x$, and $\zeta_x$ are randomly picked for each position from the range  $0 \leq \xi_x < 2\pi$, $0 \leq \theta_x \leq \pi$, and  $0 \leq \zeta_x< 2\pi$. 

\begin{figure}[ht]
\bc 
\subfigure[]{\includegraphics[width=9.0cm]{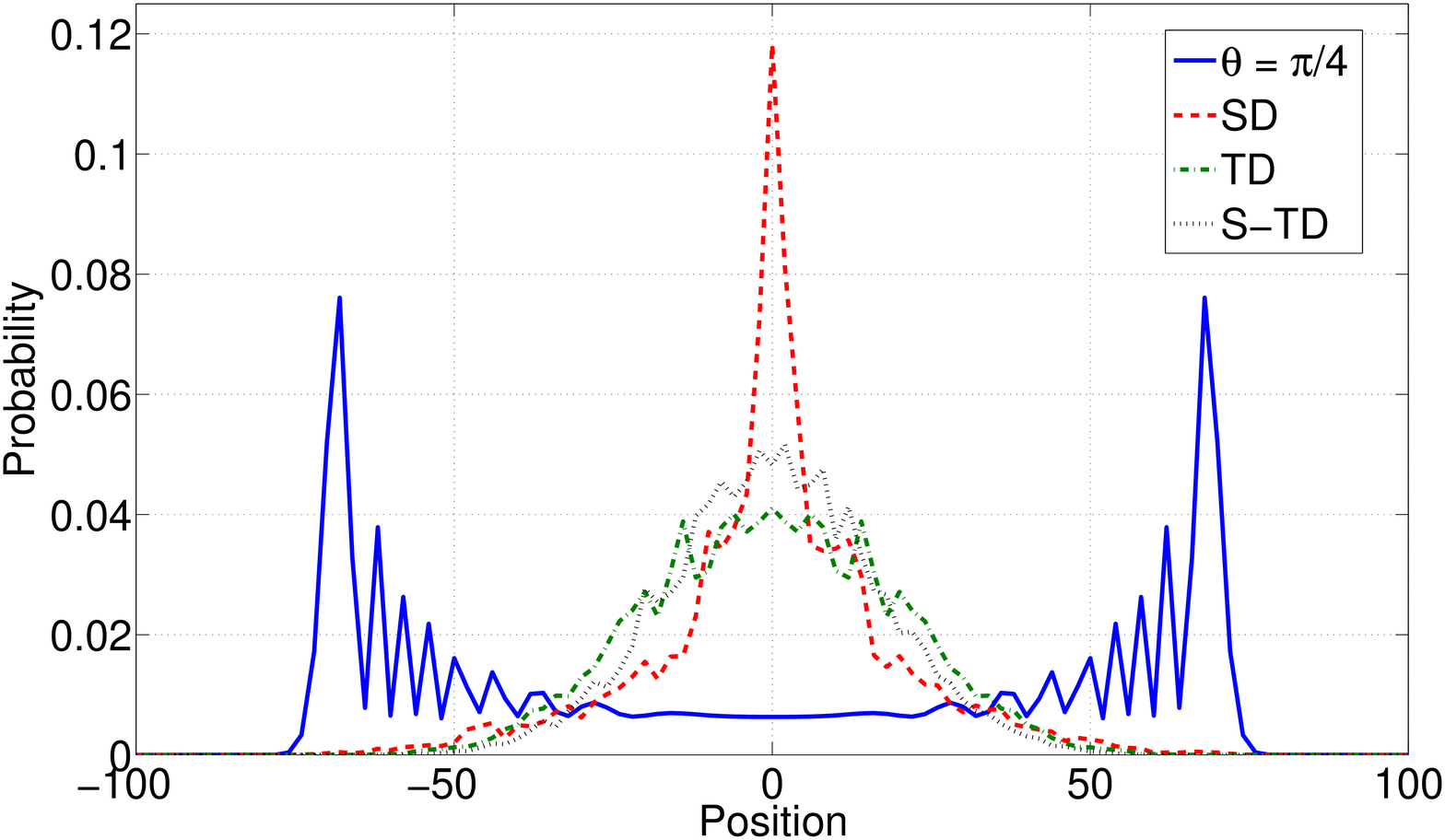} 
\label{fig:2a}}
\hskip -0.2in 
\subfigure[]{\includegraphics[width=9.0cm]{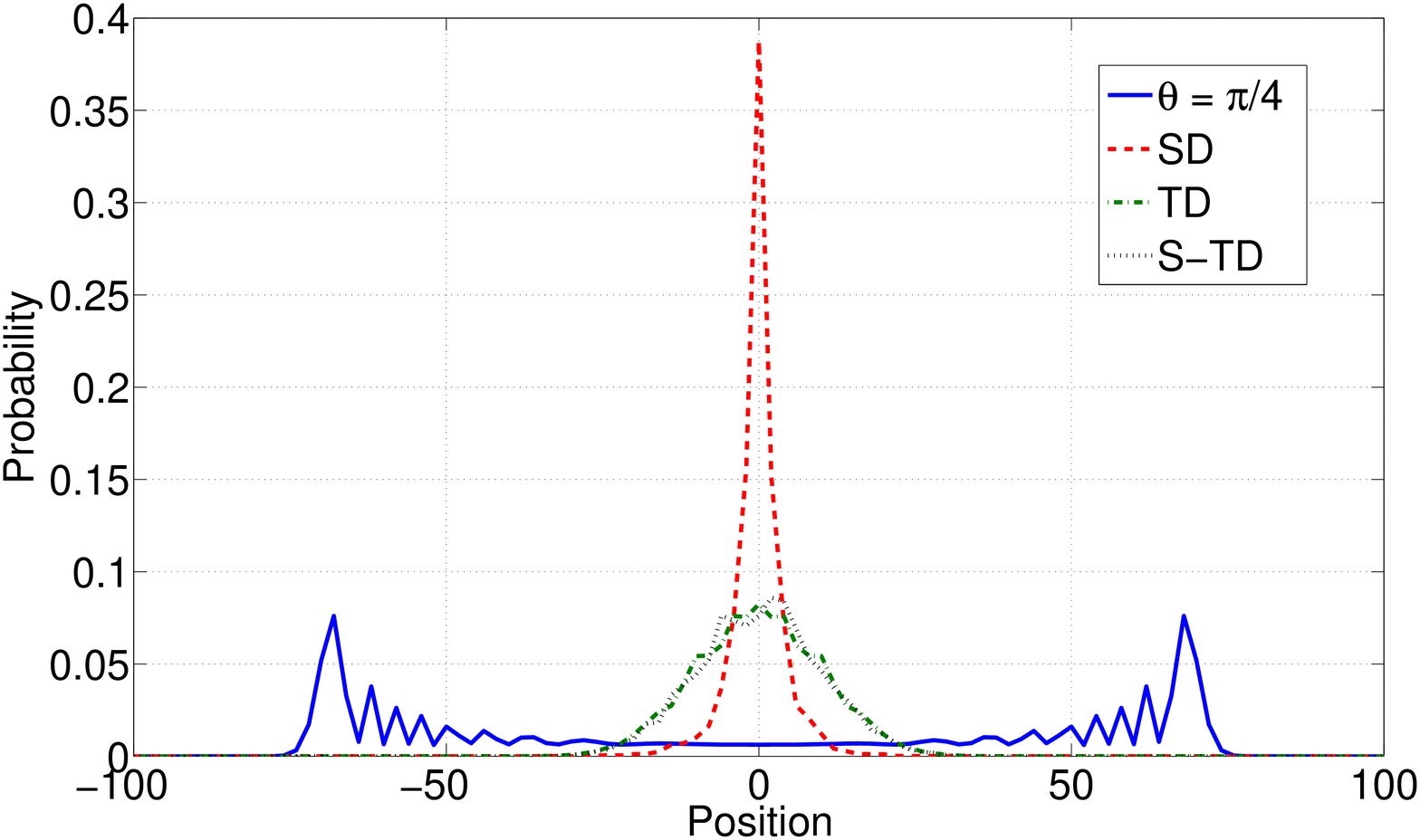}
\label{fig:2b}}
\ec
\vskip -0.7cm
\caption{\footnotesize{(Color online) Probability distribution of 1D standard DTQW  and for evolutions with spatial, temporal and spatio-temporal disorder after $t=100$ when the percentage of disorder in (a) and (b) is $20\%$  and $100\%$, respectively.  Spatial disorder shows a relatively stronger localization effect than the temporal and spatio-temporal disorder. The initial state of the particle is 
$\frac{1}{\sqrt{2}} [|\downarrow \rangle + i |\uparrow \rangle ]$.  \label{fig:2}}}
\end{figure}

The state after time $t$  with spatial disorder using single parameter position dependent coin operation $B(\theta_x)$  will be, 
\be
|\Psi_t \rangle_S = \left [ W_x(\tilde{\theta}_{x})\right ]^t |\Psi_{\rm in}\rangle = \left [   B (\tilde{\theta}_x) S_x \right ]^t |\Psi_{\rm in}\rangle.
\ee
Here $B(\tilde{\theta}_x) = \sum_x [B(\theta_x) \otimes | x \rangle \langle x|]$.
The iterative form of the state of the particle at each position $x$ and time $t+1$ in the form of the left- moving ($\psi^{\uparrow}$) and right- moving ($\psi^{\downarrow}$) components for evolution with spatial disorder are, 
\bea
\label{eq:spdis}
\begin{bmatrix}   \begin{array}{clr}
 \psi^{\uparrow}_{x, t +1}  \\
\psi^{\downarrow}_{x, t +1} 
\end{array} \end{bmatrix}
  =  \begin{bmatrix} \begin{array}{clcr}
  \mbox{~}\cos(\theta_x)      &     &     \sin(\theta_x)
  \\ -\sin(\theta_x) & &  \cos(\theta_x) 
\end{array} \end{bmatrix} \begin{bmatrix}   \begin{array}{clr}
 \psi^{\uparrow}_{x+1, t }  \\
\psi^{\downarrow}_{x-1, t } 
\end{array} \end{bmatrix}.
\eea
Decoupling the basis states, $\psi^{\uparrow}$ and $\psi^{\downarrow}$ we obtain 
\bea
\label{sddecopA}
\psi^{\uparrow}_{x, t+1} + \sin(\theta_x)\csc(\theta_{x-1}) \psi^{\uparrow}_{x, t-1}  ~~~~~~~~~~~~~~\nonumber \\
= \cot(\theta_{x-1})\sin(\theta_x) \psi^{\uparrow}_{x-1, t} + \cos(\theta_x)\psi^{\uparrow}_{x+1, t}~~~;
\eea
\bea
\label{sddecopB}
\psi^{\downarrow}_{x, t+1} + \sin(\theta_x)\csc(\theta_{x+1}) \psi^{\downarrow}_{x, t-1} ~~~~~~~~~~~~~~ \nonumber \\
= \cot(\theta_{x+1})\sin(\theta_x) \psi^{\downarrow}_{x+1, t} + \cos(\theta_x)\psi^{\downarrow}_{x-1, t}~~.
\eea
The partial differential equation form for each position $x$ with the dependency on the coin parameter $\theta_{x-1}$ for $\psi^{\uparrow}$ and
$\theta_{x+1}$ for $\psi^{\downarrow}$ is  (see Appendix (\ref{AppendixA})),
\bea 
\label{SDde}
\Bigg [\frac{\partial^2 }{\partial t^2} - \cos(\theta_x) \frac{\partial^2}{\partial x^2}+ \Big [ 1-\sin(\theta_x) \csc(\theta_{x \pm 1})\Big ] \frac{\partial}{\partial t}  +~~~~~ \nonumber \\
 \Big [ \cot(\theta_{x \pm 1})\sin(\theta_x)  - \cos(\theta_x) \Big ]  \frac{\partial}{\partial x}   
 + \Big [ 1   - \cos(\theta_x) ~~~~  \nonumber \\
+\sin(\theta_x)[\csc(\theta_{x\pm 1})    - \cot( \theta_{x\pm 1})]\Big]  \Bigg ] \psi^{\uparrow (\downarrow)}_{x, t} = 0.~~~
\eea
For solution of the form  $e^{i(k x - \omega t)}$, the dispersion relation for the equation governing the spatial disordered evolution will be
\bea
\label{disSD}
-\omega^2 + k^2 \cos(\theta_x) - i \omega \Big [ 1-\sin(\theta_x) \csc(\theta_{x \pm 1})\Big ] ~~~~~~~~~~~~ \nonumber \\
+ i k \Big [ \cot(\theta_{x \pm 1})\sin(\theta_x)  
- \cos(\theta_x) \Big ]    +  {\mathfrak A}(\theta, x)  = 0.~~
\eea
Here
\be
{\mathfrak A}(\theta, x) = 1+\sin(\theta_x)\Big [ \csc(\theta_{x\pm 1}) - \cot( \theta_{x\pm 1}) \Big ] - \cos(\theta_x).
\ee
Solving for $\omega$ by considering only the real terms in the Eq.\,(\ref{disSD}) we obtain,
\bea
\omega(k, \theta_x)  = \pm \sqrt{k^2 \cos(\theta_x) + {\mathfrak A}(\theta, x)},
\eea 
and the group velocity will be
\bea
v_g(k, \theta_x) = \frac{d \omega(k, \theta_x)}{d k}  = \pm \frac{k \cos(\theta_{x})}{\sqrt{k^2 \cos(\theta_x) + {\mathfrak A}(\theta, x)}}.
\eea
Since the coin operation is not uniform for the entire position space, the effective group velocity for the evolution with spatial disorder will be an average of the group velocity for each position $x$ (for $t$ time steps we need a position space of dimension $(2t+1)$). That is,
\bea
v_g^{SD}(k, \theta_x, t) = \sum_{x = -t}^{t} \frac{v_g(k, \theta_x)}{2t+1} .
\eea
For any random value of coin parameter in the range: $0 \leq \theta_x \leq \pi$, the  group velocity $v_g^{SD}$ will be proportional to the summation of the random values in the range: $-1 \leq  v_g(k, \theta_x)  \leq 1$.  Therefore, $|v_g^{SD}(k, \theta_x, t)|$ will be a small non-zero value when $t$ is small resulting in a initial spread of the wavepacket. But with increase in $t$ we obtain, 
\bea
v_g^{SD}(k, \theta_x, t) \approx 0,
\eea
inducing the localization in the position space.  The small spread of the wavepacket in position space before $v_g^{SD} \approx 0$  (see Fig.\,\ref{fig:2}) will be,
\bea
\label{sdvg}
D^{SD}(t) = \sum_0^{t} v_g^{SD}(k, \theta_x, t).
\eea
\par
Similarly, even for the position dependent SU(2) coin operation, the group velocity for each position will be in the range:  $-1\leq  v_g(k, \theta_x, \phi_x) \leq 1 $ (see Appendix\,\ref{AppendixB} for the complete derivation).   Therefore, with increase in $t$ we obtain  
\bea
v_g^{SD}(k, \theta_x, \phi_x, t)  = \sum_{x = -t}^{t} \frac{v_g(k, \theta_x, \phi_x)}{2t+1} \approx 0.
\eea
Thus, in general the spatial disorder in DTQW evolution induces a strong localization of the particle in position space with increase in $t$. 
\begin{figure}[ht]
\bc 
\subfigure[]{\includegraphics[width=9.0cm]{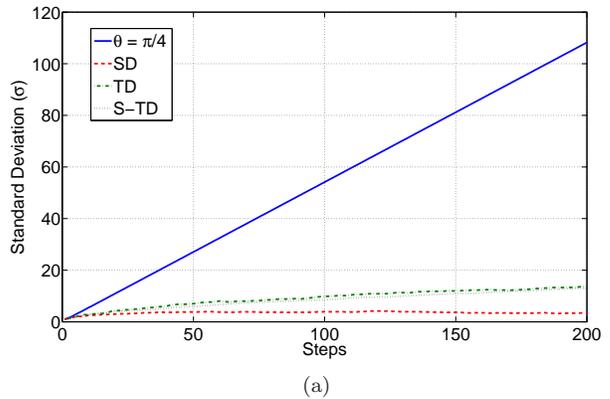} 
\label{fig:3a}}
\hskip -0.2in 
\subfigure[]{\includegraphics[width=9.0cm]{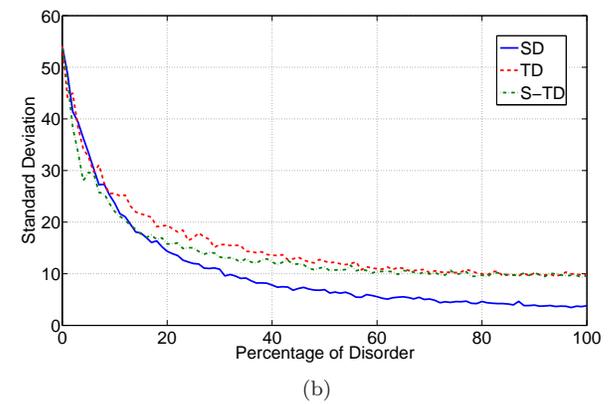}
\label{fig:3b}}
\ec
\vskip -0.7cm
\caption{\footnotesize{(Color online) (a) Standard deviation as a function of step ($t$) for normal evolution with $\theta = \pi/4$ and for evolutions with spatial, temporal and spatio-temporal disorder. (b) Standard deviation as a function of percentage of disordered operation starting from the normal evolution with $\theta = \pi/4$ after 100 steps. For evolution with disorder, standard deviation does not show any noticeable increment with increase in steps showing the signature of localization and a significant decrease in standard deviation is seen with increase in percentage of disorder.}\label{fig:3}}
\end{figure}
\subsection{Temporal disorder}
\label{td}
We define the temporal disorder in the DTQW evolution\,\cite{JM10, Cha11a} by introducing a time dependent coin operation $B(\theta_t)$  or $B (\xi_t, \theta_t, \zeta_t)$, where  $\xi_t$, $\theta_t$, and $\zeta_t$ are randomly picked for each instant of time from the range:  $0 \leq \xi_t < 2\pi$, $0 \leq \theta_t \leq \pi$, and  $0 \leq \zeta_t < 2\pi$. 
\par
The state after $t$ step with temporal disorder using a single parameter time dependent coin operation will be, 
\be
|\Psi_t \rangle_T = W_x(\theta_{t})\dots W_x(\theta_{3})W_x(\theta_{2})W_x(\theta_{1})|\Psi_{\rm in}\rangle.
\ee
The iterative form of the  state of the particle at each position $x$ and time $(t+1)$ will be identical to Eq.\,(\ref{eq:spdis})
and its decoupled expressions for $\psi^{\uparrow}$ and $\psi^{\downarrow}$ will be identical to Eqs.\,(\ref{sddecopA}) and \,(\ref{sddecopB}), respectively with only a replacement of $\theta_{t \pm 1}$ in place of $\theta_{x \pm 1}$. Therefore, the group velocity  will be,
\bea
v_g(k, \theta_t) = \frac{d \omega(k, \theta_t)}{d k}  =  \pm \frac{k \cos(\theta_{t})}{\sqrt{k^2 \cos(\theta_t) + {\mathfrak A}(\theta, t)}}.
\eea
Here 
\be
{\mathfrak A}(\theta, t) = 1+\sin(\theta_t)\Big [ \csc(\theta_{t\pm 1}) - \cot( \theta_{t\pm 1}) \Big ]
- \cos(\theta_t).
\ee
For each instance of time $t$ the group velocity will be a random value in the range: $-1 \leq v_g(k, \theta_t) \leq 1$. 
The effective group velocity for the evolution with temporal disorder will be an average of the group velocity for each instance of time $t$. That is,
\bea
v_g^{TD}(k, \theta_t, t) = \frac{1}{t} \sum_{t = 0}^{t} v_g(k, \theta_t),
\eea
and with increase in $t$ we obtain
\bea
v_g^{TD}(k, \theta_t, t) \approx 0,
\eea
resulting in the localization in position space. A small spread of the wavepacket in position space before $v_g^{TD} \approx 0$  will be, 
\bea
\label{tdvg}
D^{TD}(t) = \sum_0^{t} v_g(k, \theta_t).
\eea
Comparing Eqs.\,(\ref{sdvg}) and \,(\ref{tdvg}) we can see that Eq.\,(\ref{sdvg}) is a summation of terms with $2t+1$ term in the denominator resulting in small spread of the wavepacket with spatial disorder compared to the temporal disorder. We can see this effect in Fig.\,\ref{fig:2}. 
\par
Similarly,  with increase in $t$ for the time dependent SU(2) coin operation  we can obtain  
\bea
v_g^{TD}(k, \theta_t, \phi_t)  = \frac{1}{t} \sum_{t = 0}^{t} v_g(k, \theta_t, \phi_t, t) \approx 0.
\eea
Thus, the temporal disorder in DTQW evolution also induces a localization of the particle in position space with increase in $t$.  

\subsection{Spatio-Temporal disorder}

We define the spatio-temporal disorder in the DTQW evolution by introducing a position and time dependent coin operation $B(\theta_{x, t})$  or $B (\xi_{x,t}, \theta_{x,t}, \zeta_{x,t})$, where  $\xi_{x, t}$, $\theta_{x, t}$, and $\zeta_{x, t}$ are randomly picked for each instant of time from the range  $0 \leq \xi_{x, t} < 2\pi$, $0 \leq \theta_{x, t} \leq \pi$, and  $0 \leq \zeta_{x, t} < 2\pi$. 
\par
The state after $t$ step with spatio-temporal disorder evolution with single parameter coin operation will be  
\bea
|\Psi_t \rangle_{S-T} = \left [  B (\tilde{\theta}_{x, t}) S_x \right ]  \cdot  \cdot \left [  B (\tilde{\theta}_{x, 2}) S_x  \right ] \left [  B (\tilde{\theta}_{x, 1}) S_x  \right ] |\Psi_{\rm in}\rangle,~
\eea
where for each instance of time $t$, $B(\tilde{\theta}_{x, t}) = \sum_x [B(\theta_{x, t}) \otimes |x\rangle \langle x|]$ is the position dependent coin operation. 
The iterative form of the  state of the particle at each position $x$ and time $(t+1)$ will be identical to Eq.\,(\ref{eq:spdis})
and its decoupled expressions for $\psi^{\uparrow}$ and $\psi^{\downarrow}$ will be identical to Eqs.\,(\ref{sddecopA}) and \,(\ref{sddecopB}), respectively with only a replacement of $\theta_{x \pm 1, t \pm 1}$ in place of $\theta_{x \pm 1}$. Since the coin operation is not uniform for the entire position space and time $t$, the group velocity for the evolution with spatio-temporal disorder will be an average of the group velocity for each position $x$ and time $t$. That is, 
\bea
\label{stdvg}
v_g^{S-TD}(k, \theta_{x, t}, t) = \frac{1}{t} \sum_{t=0}^{t}\Bigg [  \sum_{x = -t}^{t} \frac{v_g(k, \theta_{x, t})}{2t+1}  \Bigg ]~.~
\eea
The  group velocity $v_g^{S-TD}$ will be proportional to the summation of the random values in the range: $-1 \leq v_g(k, \theta_{x, t})  \leq 1$.  Therefore, with increase in $t$ (size of the position space) we obtain,
\bea
v_g^{S-TD}(k, \theta_{x,t}, t) \approx 0,
\eea
inducing localization with spatio-temporal disorder. Similarly, with increase in time $t$ for the position and time dependent SU(2) coin operation we obtain 
\bea
v_g^{S-TD}(k, \theta_{x,t}, \phi_{x,t})  = \frac{1}{t} \sum_{t = 0}^{t} \Bigg [ \sum_{x=t}^{t} \frac{v_g(k, \theta_{x,t}, \phi_{x, t})}{2t+1} \Bigg ] \approx 0.~~
\eea
Thus, spatio-temporal disorder in DTQW also induces localization of the particle in position space with increase in $t$. 
\par
In Fig.\ref{fig:2}, we show the probability distribution after 100 steps of the standard DTQW evolution  and the evolution with spatial, temporal and spatio-temporal disordered operations. In Fig.\,\ref{fig:2a}, only 20\% of the coin operations ($\theta_x$, $\theta_t$, and $\theta_{x, t}$) are randomly chosen for the evolution with remaining 80\% of the operations with $\theta =\pi/4$. 
Even a small fraction of disorder shows the trend towards localization and that can be seen as a weak localization regime.  In Fig.\,\ref{fig:2b}, the coin operations ($\theta_x$, $\theta_t$, and $\theta_{x, t}$) are randomly chosen for the complete evolution. Spatial disorder has a relatively stronger effect on the diffusion compared to the temporal and spatio-temporal disorder. 
\begin{figure}[ht]
\bc 
\subfigure[SD]{\includegraphics[width=8.2cm]{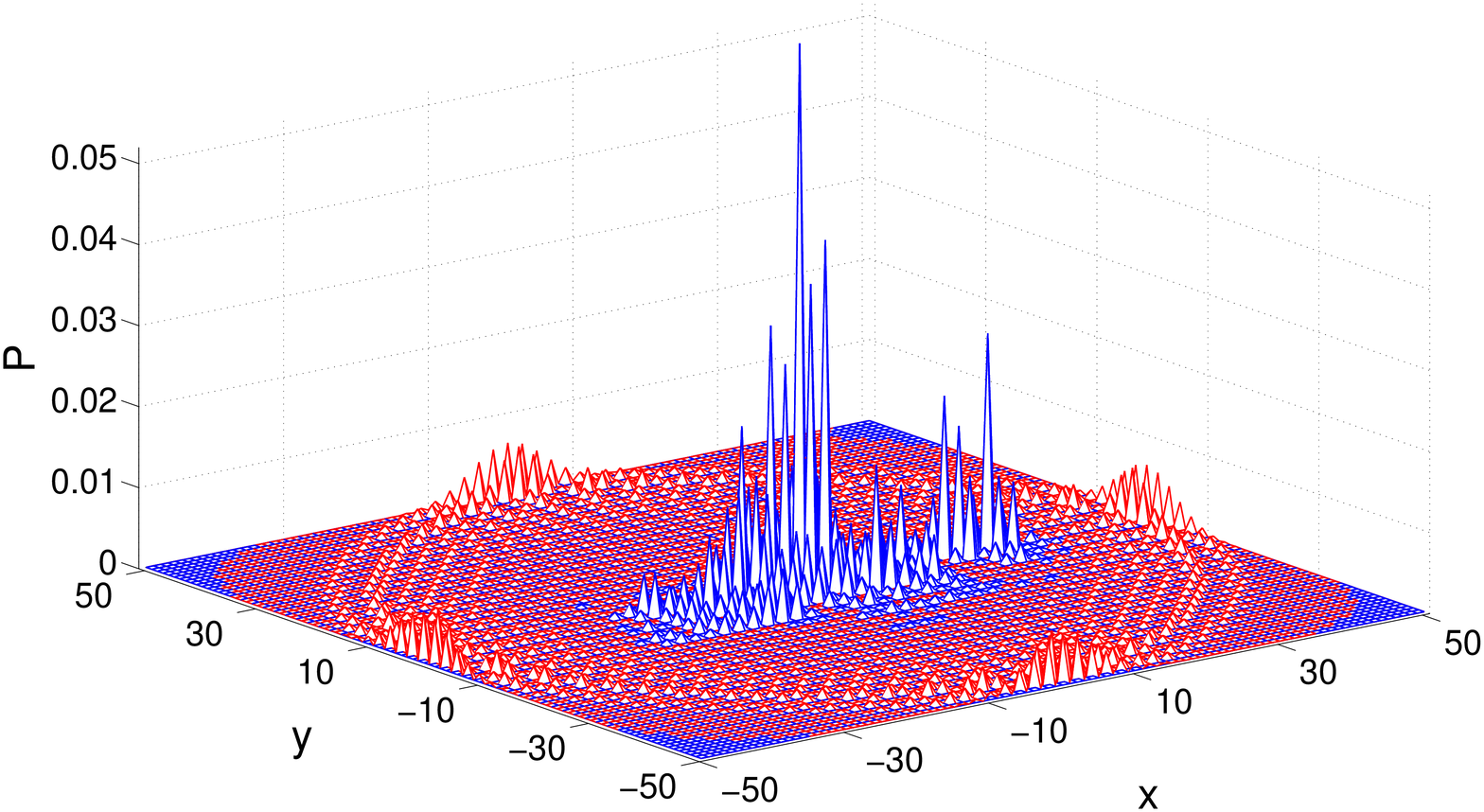} 
\label{fig:4a}}
\hskip -0.2in 
\subfigure[TD]{\includegraphics[width=8.2cm]{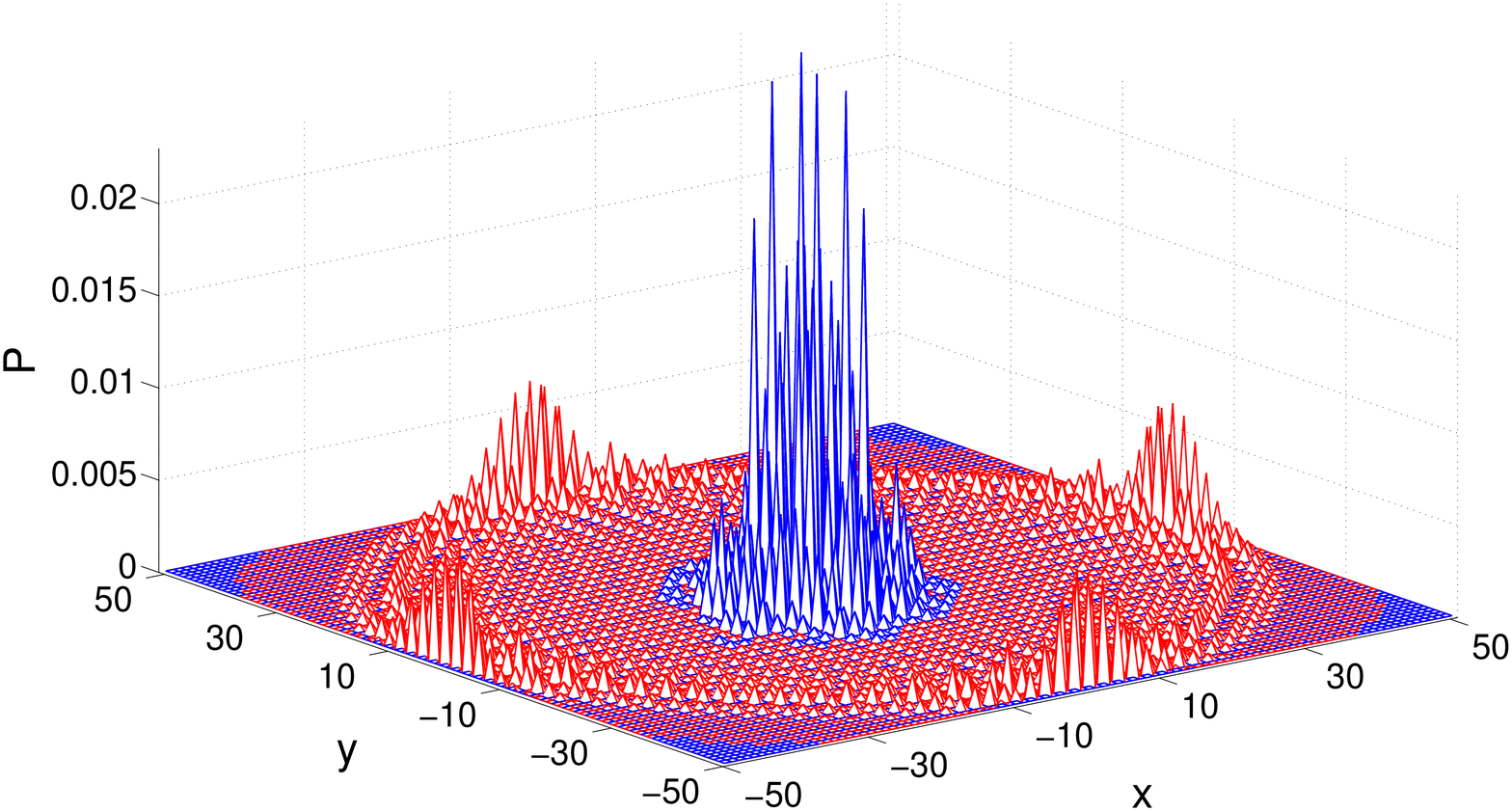}
\label{fig:4b}}\\
\subfigure[S-TD]{\includegraphics[width=8.2cm]{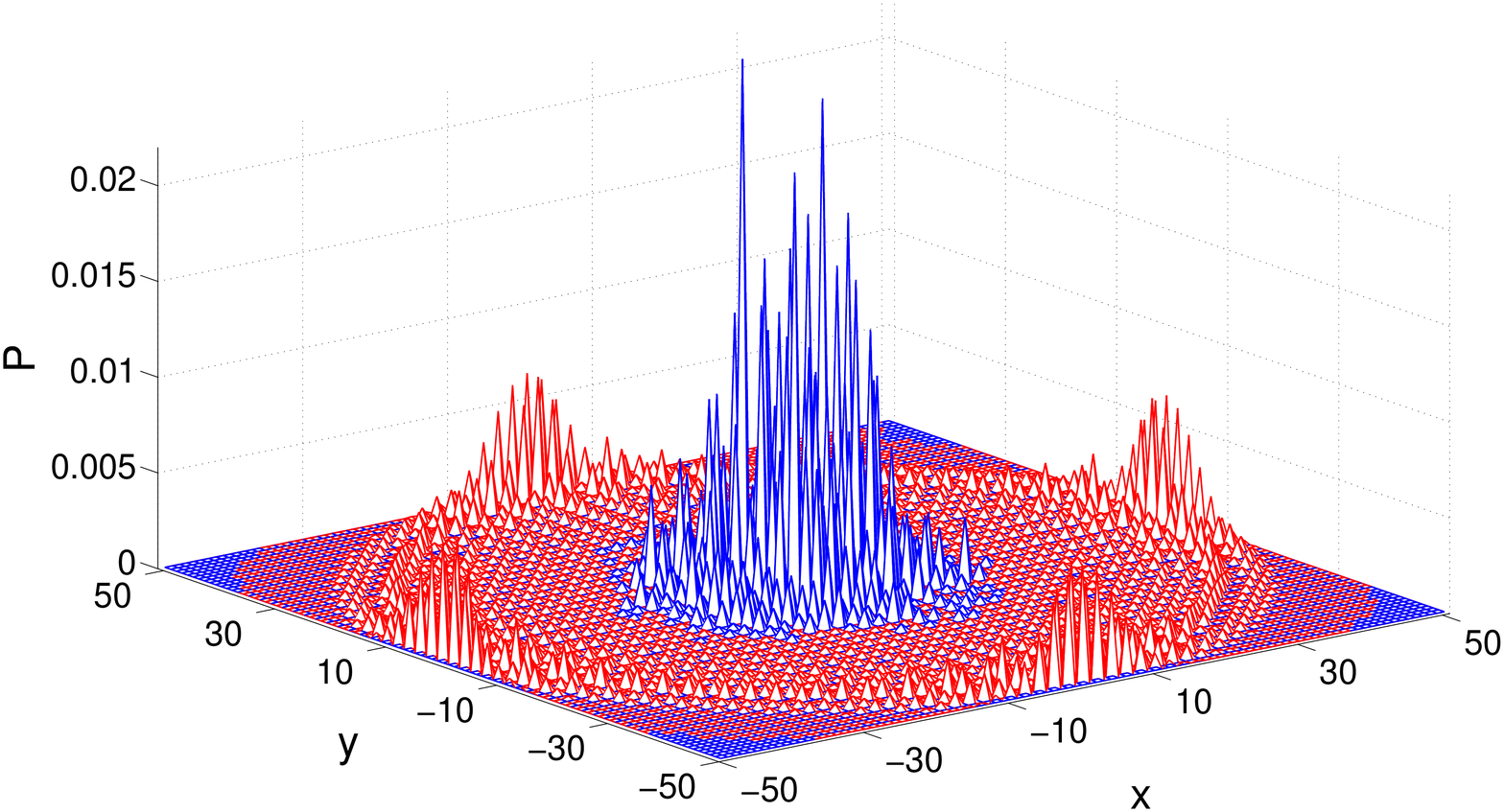}
\label{fig:4c}}
\ec
\vskip -0.7cm
\caption{\footnotesize{(Color online) Comparison of the probability distribution after 50 steps of 2D standard DTQW evolution with (a) spatial disorder, (b) temporal, and (c) spatio-temporal disorder. We can see the localization of distribution for disordered evolution when compared with the standard evolution.}\label{fig:4}}
\end{figure}
\par
In Fig.\,\ref{fig:3}, we show the standard deviation obtained numerically for the DTQW with standard ($\theta =\pi/4$),  spatial, temporal and spatio-temporal evolutions. Fig.\,\ref{fig:3a}, shows a significant increase in standard deviation as a function of steps ($t$) for standard DTQW evolution but for evolution with disorder, absence of noticeable increment with disorder shows the signature of strong localization. In Fig.\,\ref{fig:3b}, standard deviation after 100 steps of walk as a function of percentage of disorder is shown. We can see a sudden decrease in standard deviation even for small increase in percentage of disorder. In all this,  localization due to spatial disorder is stronger compared to the temporal and spatio-temporal disorder.   

\section{Localization in two-dimension }
\label{2dLoc}
Two-state DTQW on a square lattice can be realized by evolving the particle in one axis followed by the evolution in the other axis\,\cite{FB11}. Alternatively, a two state walk on square lattice can also be constructed using different Pauli basis sates as translational states for the two axis\,\cite{CSB10, Cha11b} and we will use this scheme to study  the effect of disorder in two-dimension. For convenience we choose $\{|\uparrow \rangle$,  $|\downarrow \rangle \}$ and $\{ |+\rangle = \frac{1}{\sqrt 2}(|\uparrow \rangle + |\downarrow \rangle), |-\rangle = \frac{1}{\sqrt 2}(|\uparrow \rangle - |\downarrow \rangle)\}$, eigenstates of the Pauli operators $\sigma_3= \begin{bmatrix} 1 & ~~0\\ 0 & -1 \end{bmatrix}$ and $\sigma_1 = \begin{bmatrix} 0 &
  1\\ 1 & 0 \end{bmatrix}$ as basis states for $x-$ and $y-$axis, respectively. 
Because of the use of different Pauli basis for translation in different axis, DTQW in 2D can be realized without using a quantum coin operation.  That is, each step of DTQW in 2D composes of  $S_{0, y}S_{x, 0}$, where  
 \bea
 S_{x, 0}  \equiv   \sum_{x, y}  \Big [  |\uparrow \rangle\langle \uparrow |\otimes|x-1, y \rangle\langle   x, y |  \nonumber   \\
 + | \downarrow \rangle\langle \downarrow |\otimes |x +1, y \rangle\langle x, y| \Big ] \\
 S_{y, 0} \equiv   \sum_{x, y}  \Big [ |+ \rangle\langle +|\otimes|x, y-1\rangle\langle x, y|    \nonumber   \\
 + |- \rangle\langle-|\otimes |x, y+1\rangle\langle x, y| \Big ].
 \eea
 However, quantum coin operation is a resource that can be exploited to control the probability distribution.  Therefore, we will introduce a quantum coin operation after evolution in each direction and the state after $t$ step will be
\be
|\Psi^{2d}_{t}\rangle = \Bigg[ \Big [ B_{x}(\theta)\otimes  {\mathbbm 1} \Big ] S_{0, y} \Big [ B_{y}(\vartheta)\otimes  {\mathbbm 1} \Big ]  S_{x, 0} \Bigg ] ^t|\Psi_{in}\rangle.
\ee
Here $B_{x}(\theta) = B(\theta)$, same as coin operation used for 1D DTQW evolution [Eq.\,(\ref{coinOP})] and   
\bea
B_{y}(\vartheta)  = \cos (\vartheta) |+ \rangle \langle + | + \sin (\vartheta) |+ \rangle \langle -|   \nonumber \\
 - \sin (\vartheta) |- \rangle \langle + | + \cos (\vartheta) |- \rangle \langle - |.
\eea
\begin{figure}[ht]
\bc 
\subfigure[]{\includegraphics[width=8.2cm]{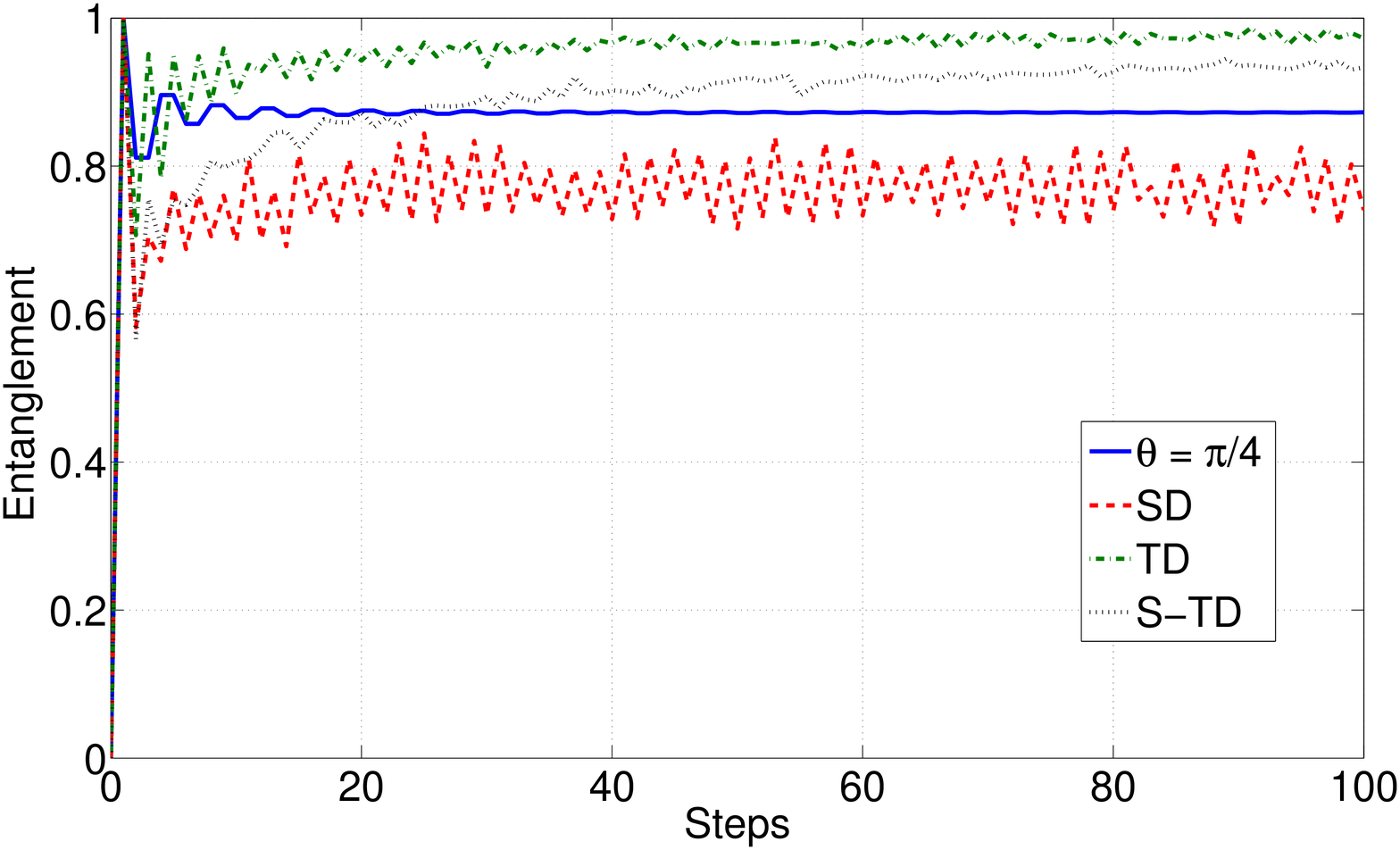} 
\label{fig:5a}}
\hskip -0.2in 
\subfigure[]{\includegraphics[width=8.2cm]{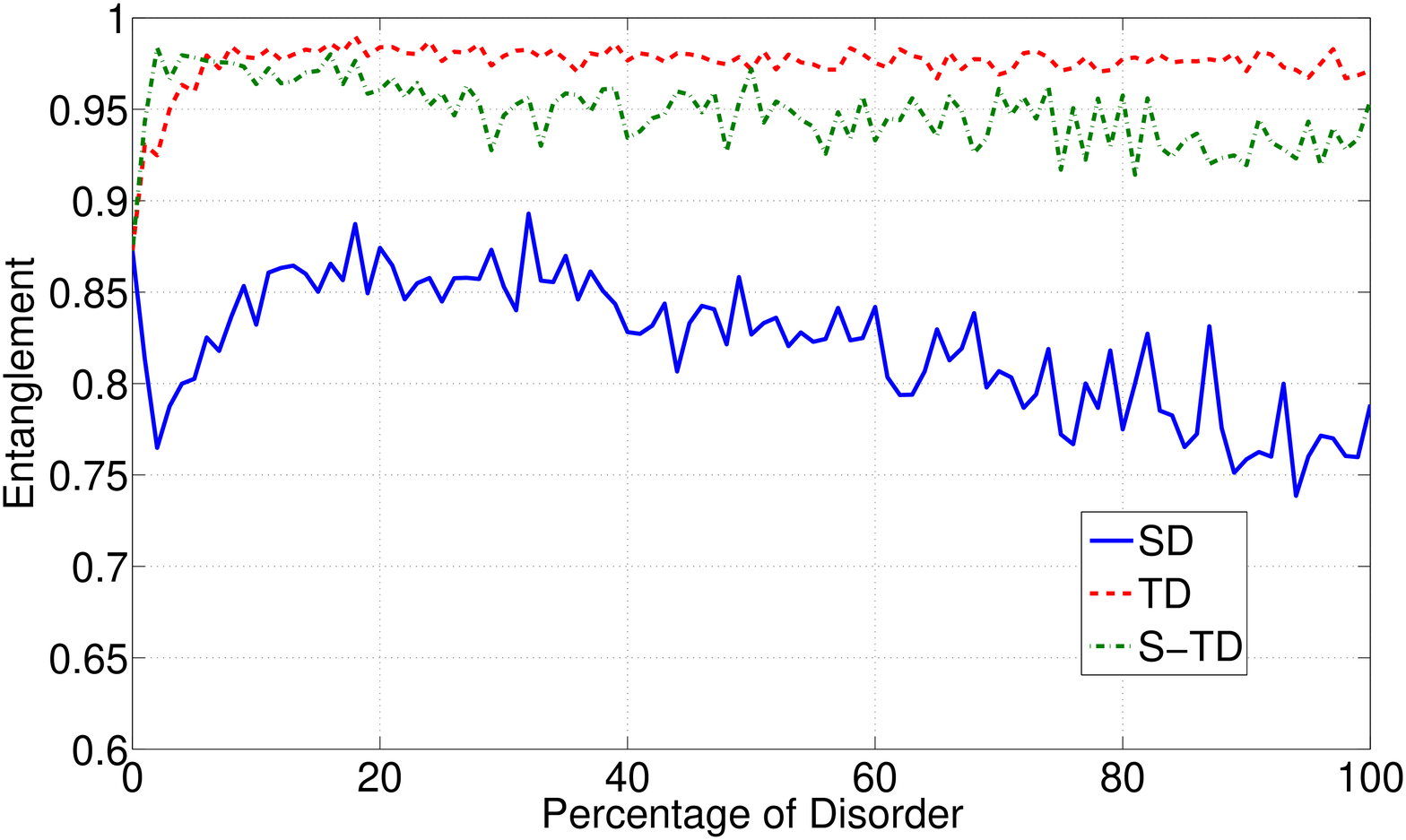}
\label{fig:5b}}
\ec
\vskip -0.7cm
\caption{\footnotesize{(Color online) (a)  Degree of entanglement as a function of steps and (b) as a function of percentage of disorder (after 100 step) using von Neumann entropy as a measure  for standard evolution of 1D DTQW and evolution with spatial, temporal and spatio-temporal disordered coin operations.  Spatial disorder suppresses the degree of entanglement by a small amount whereas, temporal and spatio-temporal disorder enhances the entanglement.}
\label{fig:5}}
\end{figure}
The $\psi^{\uparrow}$ and $\psi^{\downarrow}$ components at any instance of time are dependent on the states at four different positions from the previous instance of time (see Appendix\,\ref{AppendixC} for the iterative form of the evolution).  In the same way as we introduced disorder into the 1D DTQW evolution, we can introduce the 
spatial, temporal and spatio-temporal disorder, by randomly choosing different $0 \le \theta \leq \pi $ and $0 \le \vartheta \leq \pi $ for each position, time and both position-time, respectively. The numerical evolution using different form of disordered coin operation in 2D DTQW is shown in Fig.\,\ref{fig:4}. Although interference effect is seen in the distribution, we note that the localization of the probability distribution is observed around the origin with all the three forms of disorder.  This localization seen in 2D using two-state particle is due to the dynamical disorder unlike the localization caused by a special choice of coin and initial state in Grover walk using four-state particle \cite{IKK04}. 
\begin{figure}[ht]
\bc 
\subfigure[]{\includegraphics[width=8.2cm]{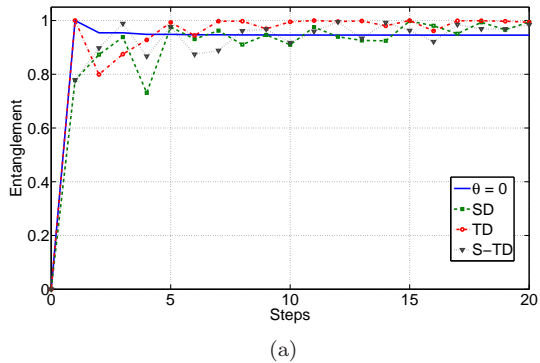}
\label{fig:6}}
\ec
\vskip -0.7cm
\caption{\footnotesize{(Color online)  Degree of entanglement as a function of steps ($t$) using von Neumann entropy as a measure  for standard evolution and evolution with spatial, temporal and spatio-temporal disorder. We can see a small enhancement of entanglement for disordered evolution.}\label{fig:6}}
\end{figure}

\section{Disorder and effect on the entanglement between the particle and position space}
\label{entDis}
Entanglement, is a very useful resource in quantum information processing and in an entangled system, noise or disorder results in decrease or death of entanglement.
But in our study the disorder introduced does not affect the unitarity of the evolution. Therefore, its effect on entanglement is not significantly large. Interesting we note an increase in entanglement between the particle and position space with temporal and spatio-temporal disorder and a small decrease for spatial disorder when compared with the DTQW evolution using $\theta = \pi/4$.
\par
Entanglement measure between the particle and position space in DTQW is one of the simpler way to judge the presence of the particle in superposition of the position space.  In Fig.\,\ref{fig:3} we show the effect of disorder on the entropy of entanglement between the particle and the position space measured using von Neumann entropy, 
\be
E(\rho) = S(\rho_c) = -{\rm Tr} [\rho_c \log_2( \rho_c)],
\ee
where $\rho_c$  is the density matrix of the particle space after tracing out the position space. That is, for one dimensional walk $\rho_c =  Tr_p (\rho)  =  \sum_x \langle x |( |\Psi_t \rangle \langle \Psi_t |)| x \rangle$ with subscript $c$ and $p$ representing the particle and position Hilbert space, respectively. In Fig.\,\ref{fig:5a}, degree of entanglement as a function of time is shown and it is directly proportional to the size of the position space in which the particle is found in superposition. Because of strong localization induced by spatial disorder the sites in which particle is seen in superposition is relatively less when compared to the DTQW with $\theta = \pi/4$ resulting in the decrease of entanglement. Though $\theta= \pi/4$ spreads the distribution wide in position space, but the amplitude of the particle is largely concentrated on a small number of position space which are at the extreme points in the position space. Therefore, even with localization around the origin the spread of amplitude in superposition of  position space with temporal and spatio-temporal disorder is large enhancing the entanglement. In Fig.\,\ref{fig:5b}, we show the degree of entanglement with increase in percentage of disorder and a similar effect of decrease in entanglement with spatial disorder and enhancement is temporal and spatio-temporal disorder is seen.
\par
Disorder on square lattice has a very small effect on the degree of entanglement when compared to the effect it had on the one dimensional DTQW system. However small it is, we can still note the enhancement of entanglement with all the forms of disorder in two-dimension (see Fig.\,\ref{fig:6}).

\section{Conclusion} 
\label{conc}
 DTQW which normally diffuse quadratically faster when compared to the classical random walk localizes with disorder. Interestingly, the interference during the evolution is what plays a significant role for both, faster diffusion and for localization. Using the dispersion relation and the group velocity of the DTQW in continuum limit we showed the dynamics of the disordered evolution leading to the localization.  Our study modeled using the spatial, temporal and spatio-temporal disordered unitary operations showed a noticeable effect on the degree of entanglement between the particle and position space. Disorder in general, enhances the entanglement in both, one- an two-dimensions for all disorder studied except for spatial disorder in one-dimension where a small suppression of entanglement is seen. Anderson localization is a universal phenomenon and in addition, it can also enhance the entanglement of the particle with the position space leaving open, both the negative and positive implications on its applications to algorithmic and to understand the dynamics in various other quantum systems. Randomized unitary operations can assist localization (storage)  of quantum state in variety of physical systems and applications in quantum information processing.   This opens up an important question for future, interplay between the disorder and entanglement and give more options to further explore topological phase, and understand physical process in nature where disorder is a natural phenomenon.
\begin{widetext}
\appendix
\section{Decoupling the two basis states in the evolution with spatial disorder and its differential form}
\label{AppendixA}

The iterative form of the state of the particle at each position $x$ and time $t=1$ with spatial disorder using a single parameter $B(\theta_x)$ are
\bea
\label{a1}
\begin{bmatrix}   \begin{array}{clr}
 \psi^{\uparrow}_{x, t +1}  \\
\psi^{\downarrow}_{x, t +1} 
\end{array} \end{bmatrix}
  =  \begin{bmatrix} \begin{array}{clcr}
  \mbox{~}\cos(\theta_x)      &     &     \sin(\theta_x)
  \\ -\sin(\theta_x) & &  \cos(\theta_x) 
\end{array} \end{bmatrix} \begin{bmatrix}   \begin{array}{clr}
 \psi^{\uparrow}_{x+1, t }  \\
\psi^{\downarrow}_{x-1, t } 
\end{array} \end{bmatrix};
\eea
\bea
\label{a2}
\psi^{\uparrow}_{x, t+1}& =& \cos(\theta_x) \psi^{\uparrow}_{x+1, t} + \sin(\theta_x) \psi^{\downarrow}_{x-1, t} ~~;\\
\label{a3}
\psi^{\downarrow}_{x, t+1} &=& -\sin(\theta_x) \psi^{\uparrow}_{x+1, t} + \cos(\theta_x) \psi^{\downarrow}_{x-1, t} .
\eea
From Eq.\,(\ref{a2}) we can write
\bea
\label{a4}
\psi^{\downarrow}_{x-1, t} &= &\frac{\psi^{\uparrow}_{x, t+1} - \cos(\theta_x)\psi^{\uparrow}_{x+1, t}}{\sin(\theta_x)}  ~~;\\
\label{a5}
\psi^{\downarrow}_{x, t+1} &=& \frac{\psi^{\uparrow}_{x+1, t+2} - \cos(\theta_{x+1})\psi^{\uparrow}_{x+2, t+1}}{\sin(\theta_{x+1})}.  
\eea
Substituting Eqs.\,(\ref{a4}) and (\ref{a5}) into Eq.\,(\ref{a3}) and simplifying we obtain
\bea
\label{a6}
\psi^{\uparrow}_{x, t+1} + \frac{\sin(\theta_x)}{\sin(\theta_{x-1})} \psi^{\uparrow}_{x, t-1}  = \frac{\cos(\theta_{x-1})\sin(\theta_x)}{\sin(\theta_{x-1})} \psi^{\uparrow}_{x-1, t} + \cos(\theta_x)\psi^{\uparrow}_{x+1, t} ;\\
\label{a7}
\psi^{\uparrow}_{x, t+1} + \sin(\theta_x)\csc(\theta_{x-1}) \psi^{\uparrow}_{x, t-1}  = \cot(\theta_{x-1})\sin(\theta_x) \psi^{\uparrow}_{x-1, t} + \cos(\theta_x)\psi^{\uparrow}_{x+1, t}
\eea 
Subtracting both sides of the Eq.\,(\ref{a7})  by
 \bea
 \label{a8}
\Big[ 1+ \sin(\theta_x) \csc(\theta_{x-1}) + \cos(\theta_x) + \cot(\theta_{x-1})\sin(\theta_x) \Big ] \psi^{\uparrow}_{x, t} - \psi^{\uparrow}_{x, t-1} - \cos(\theta_x)\psi^{\uparrow}_{x-1, t} ,
 \eea
we obtain a difference form which can be written in the differential form as 
\bea
\label{a9}
\Bigg [ \frac{\partial^2}{\partial t^2} -\cos(\theta_x)\frac{\partial^2}{\partial x^2}  + \Big [ 1-\sin(\theta_x) \csc(\theta_{x-1})\Big ] \frac{\partial}{\partial t} + \Big [ \cot(\theta_{x-1})\sin(\theta_x) - \cos(\theta_x) \Big ]  \frac{\partial}{\partial x}   \nonumber \\
 + \Big [ 1+\sin(\theta_x)\csc(\theta_{x-1}) - \cos(\theta_x) - \cot( \theta_{x-1}) \sin(\theta_x)\Big]  \Bigg ] \psi^{\uparrow}_{x, t} = 0 .
\eea
Similarly,  expression for $\psi^{\downarrow}$ will be
\bea
\label{a10}
\psi^{\downarrow}_{x, t+1} + \frac{\sin(\theta_x)}{\sin(\theta_{x+1})} \psi^{\downarrow}_{x, t-1}  = \frac{\cos(\theta_{x+1})\sin(\theta_x)}{\sin(\theta_{x+1})} \psi^{\downarrow}_{x+1, t} + \cos(\theta_x)\psi^{\downarrow}_{x-1, t} \\
\label{a11}
\psi^{\downarrow}_{x, t+1} + \sin(\theta_x)\csc(\theta_{x+1}) \psi^{\downarrow}_{x, t-1}  = \cot(\theta_{x+1})\sin(\theta_x) \psi^{\downarrow}_{x+1, t} + \cos(\theta_x)\psi^{\downarrow}_{x-1, t}.
\eea
Subtracting both sides of the Eq.\,(\ref{a11})  by
 \bea
 \label{a12}
[1+ \sin(\theta_x) \csc(\theta_{x+1}) + \cos(\theta_x) + \cot(\theta_{x+1})\sin(\theta_x) ] \psi^{\downarrow}_{x, t} - \psi^{\downarrow}_{x, t-1} - \cos(\theta_x)\psi^{\downarrow}_{x+1, t},
 \eea
we obtain a difference form which can be written in the differential form  as
\bea
\label{a13}
\Bigg [ \frac{\partial^2}{\partial t^2} -\cos(\theta_x)\frac{\partial^2}{\partial x^2}  + \Big [ 1-\sin(\theta_x) \csc(\theta_{x+1})\Big ] \frac{\partial}{\partial t} + \Big [ \cot(\theta_{x+1})\sin(\theta_x) - \cos(\theta_x) \Big ]  \frac{\partial}{\partial x}   \nonumber \\
 + \Big [ 1+\sin(\theta_x)\csc(\theta_{x+1}) - \cos(\theta_x) - \cot( \theta_{x+1}) \sin(\theta_x)\Big]  \Bigg ] \psi^{\downarrow}_{x, t} = 0. 
\eea
By replacing $\theta_{x \pm 1}$  by $\theta$  we get back  the differential form for a single parameter standard DTQW evolution, Eq.\,(\ref{KGe}).

\section{ Decoupling the two basis states, dispersion relation and group velocity for evolution with spatial disorder using SU(2) coin operation}
\label{AppendixB}

The iterative form of the state of the particle at each position $x$ and time $t=1$ with spatial disorder using a three parameter $B(\xi_x, \theta_x, \zeta_x)$ are

\bea
\label{b1}
\begin{bmatrix}   \begin{array}{clr}
 \psi^{\uparrow}_{x, t +1}  \\
\psi^{\downarrow}_{x, t +1} 
\end{array} 
\end{bmatrix}  =      \begin{bmatrix}     \begin{array}{clcr}
  \mbox{~~}e^{i(\xi_x + \zeta_x)}\cos(\theta_x)      &     &    e^{i(\xi_x - \zeta_x)} \sin(\theta_x)
  \\ -e^{-i(\xi_x -\zeta_x)}\sin(\theta_x) & &  e^{-i(\xi_x+\zeta_x)}\cos(\theta_x) 
\end{array} \end{bmatrix}     \begin{bmatrix}
 \begin{array}{clr}
 \psi^{\uparrow}_{x+1, t }  \\
\psi^{\downarrow}_{x-1, t } 
\end{array} \end{bmatrix}.
\eea
\bea
\label{b2}
\psi^{\uparrow}_{x, t+1}& =& e^{i(\xi_x + \zeta_x)}\cos(\theta_x) \psi^{\uparrow}_{x+1, t} +e^{i(\xi_x - \zeta_x)} \sin(\theta_x) \psi^{\downarrow}_{x-1, t} ;\\
\label{b3}
\psi^{\downarrow}_{x, t+1} &=& -e^{-i(\xi_x -\zeta_x)}\sin(\theta_x) \psi^{\uparrow}_{x+1, t} + e^{-i(\xi_x+\zeta_x)}\cos(\theta_x) \psi^{\downarrow}_{x-1, t} .
\eea
From Eq.\,(\ref{b2}) we can write
\bea
\label{b4}
\psi^{\downarrow}_{x-1, t} &= &\frac{\psi^{\uparrow}_{x, t+1} - e^{i(\xi_x + \zeta_x)}\cos(\theta_x)\psi^{\uparrow}_{x+1, t}}{e^{i(\xi_x - \zeta_x)}\sin(\theta_x)}  \\
\label{b5}
\psi^{\downarrow}_{x, t+1} &=& \frac{\psi^{\uparrow}_{x+1, t+2} - e^{i(\xi_{x-1} + \zeta_{x-1})}\cos(\theta_{x+1})\psi^{\uparrow}_{x+2, t+1}}{e^{i(\xi_{x+1} - \zeta_{x+1})}\sin(\theta_{x+1})}.  
\eea
Substituting Eqs.\,(\ref{b4}) and (\ref{b5}) into Eq.\,(\ref{b3}) and simplifying we obtain
\bea
\label{b6}
\psi^{\uparrow}_{x, t+1} + \Bigg [ e^{i  ( \xi_{x-1} - \zeta_{x-1} - \xi_{x}+\zeta_x   )} \sin(\theta_x)\sin(\theta_{x-1}) + e^{i  ( 3\zeta_{x-1} +  \xi_{x} - \zeta_{x} + \xi_{x-1}  )}\frac{\cos^2(\theta_{x-1}) \sin(\theta_x)}{\sin(\theta_{x-1})} \Bigg ]  \psi^{\uparrow}_{x, t-1}   \nonumber \\
= e^{i ( \xi_{x} + \zeta_{x}  )}  \cos(\theta_x)\psi^{\uparrow}_{x+1, t} + e^{i  ( 2\zeta_{x-1}+ \xi_{x} - \zeta_{x}  )} \frac{\cos(\theta_{x-1})\sin(\theta_x)}{\sin(\theta_{x-1})} \psi^{\uparrow}_{x-1, t}.
\eea 
Subtracting both sides of the Eq.\,(\ref{b6})  by
 \bea
 \label{b7}
\Bigg[ 1+ e^{i  ( \xi_{x-1} - \zeta_{x-1} - \xi_{x}+\zeta_x   )} \sin(\theta_x)\sin(\theta_{x-1}) + e^{i  ( 3\zeta_{x-1} +  \xi_{x} - \zeta_{x} + \xi_{x-1}  )}\cos(\theta_{x-1})\cot(\theta_{x-1}) \sin(\theta_x) \nonumber \\
+  e^{i  ( 2\zeta_{x-1}+ \xi_{x} - \zeta_{x}  )} \cot(\theta_{x-1})\sin(\theta_x) \Bigg ] \psi^{\uparrow}_{x, t} - \psi^{\uparrow}_{x, t-1} - e^{i(\xi_x +\zeta_x)}\cos(\theta_x)\psi^{\uparrow}_{x-1, t}~~ ,
 \eea
we obtain a difference form which can be written in the differential form  as
\bea
\label{b8}
\Bigg [ \frac{\partial^2}{\partial t^2} - e^{i(\xi_x +\zeta_x)}\cos(\theta_x)\frac{\partial^2}{\partial x^2}  + \Big [ 1- \sin(\theta_x) \Big (  e^{i  ( \xi_{x-1} - \zeta_{x-1} - \xi_{x}+\zeta_x   )} \sin(\theta_{x-1}) + e^{i  ( 3\zeta_{x-1} +  \xi_{x} - \zeta_{x} + \xi_{x-1}  )}\cos(\theta_{x-1})\cot(\theta_{x-1}) \Big ) \Big ] \frac{\partial}{\partial t}  \nonumber \\
+ \Big [ e^{i  ( 2\zeta_{x-1}+ \xi_{x} - \zeta_{x}  )} \cot(\theta_{x-1})\sin(\theta_x) - e^{i(\xi_x +\zeta_x)}\cos(\theta_x) \Big ]  \frac{\partial}{\partial x}   +  \Big [ 1+ \sin(\theta_x) \Big (  e^{i  ( \xi_{x-1} - \zeta_{x-1} - \xi_{x}+\zeta_x   )} \sin(\theta_{x-1})  \nonumber \\
+ e^{i  ( 3\zeta_{x-1} +  \xi_{x} - \zeta_{x} + \xi_{x-1}  )}\cos(\theta_{x-1})\cot(\theta_{x-1}) \Big )  
 -    e^{i  ( 2\zeta_{x-1}+ \xi_{x} - \zeta_{x}  )} \cot(\theta_{x-1})\sin(\theta_x) - e^{i(\xi_x +\zeta_x)}\cos(\theta_x)  \Big ]   \Bigg ] \psi^{\uparrow}_{x, t} = 0.~~~ 
\eea
For solution of the form  $e^{i(k x - \omega t)}$, the dispersion relation will be
\bea
\label{b9}
-\omega^2 + e^{i(\xi_x +\zeta_x)}\cos(\theta_x)(k^2)  - \Big [ 1- \sin(\theta_x) \Big (  e^{i  ( \xi_{x-1} - \zeta_{x-1} - \xi_{x}+\zeta_x   )} \sin(\theta_{x-1}) + e^{i  ( 3\zeta_{x-1} +  \xi_{x} - \zeta_{x} + \xi_{x-1}  )}\cos(\theta_{x-1})\cot(\theta_{x-1}) \Big ) \Big ] (i\omega)  \nonumber \\
+ \Big [ e^{i  ( 2\zeta_{x-1}+ \xi_{x} - \zeta_{x}  )} \cot(\theta_{x-1})\sin(\theta_x) - e^{i(\xi_x +\zeta_x)}\cos(\theta_x) \Big ]  (ik)   +  \Big [ 1+ \sin(\theta_x) \Big (  e^{i  ( \xi_{x-1} - \zeta_{x-1} - \xi_{x}+\zeta_x   )} \sin(\theta_{x-1})  \nonumber \\
+ e^{i  ( 3\zeta_{x-1} +  \xi_{x} - \zeta_{x} + \xi_{x-1}  )}\cos(\theta_{x-1})\cot(\theta_{x-1}) \Big )  
 -    e^{i  ( 2\zeta_{x-1}+ \xi_{x} - \zeta_{x}  )} \cot(\theta_{x-1})\sin(\theta_x) - e^{i(\xi_x +\zeta_x)}\cos(\theta_x)  \Big ]  = 0.~~~ 
\eea
Considering only the real part of Eq.(\ref{b9}) we obtain,
\bea
\label{b10}
-\omega^2 - \Big [  \sin(\theta_x) \Big (  \sin( \xi_{x-1} - \zeta_{x-1} - \xi_{x}+\zeta_x ) \sin(\theta_{x-1})
 + \sin(3\zeta_{x-1} +  \xi_{x} - \zeta_{x} + \xi_{x-1} )\cos(\theta_{x-1})\cot(\theta_{x-1}) \Big ) \Big ] \omega  \nonumber \\
+ \Big[\cos(\xi_x +\zeta_x)\cos(\theta_x)\Big ] k^2  + \Big [-\sin ( 2\zeta_{x-1}+ \xi_{x} - \zeta_{x} ) \cot(\theta_{x-1})\sin(\theta_x) + \sin(\xi_x +\zeta_x)\cos(\theta_x) \Big ] k   \nonumber \\
 +  \Big [ 1+ \sin(\theta_x) \Big (  \cos(\xi_{x-1} - \zeta_{x-1} - \xi_{x}+\zeta_x   ) \sin(\theta_{x-1})  
+\cos(3\zeta_{x-1} +  \xi_{x} - \zeta_{x} + \xi_{x-1}  )\cos(\theta_{x-1})\cot(\theta_{x-1}) \Big )  
\nonumber \\
 -   \cos( 2\zeta_{x-1}+ \xi_{x} - \zeta_{x}  ) \cot(\theta_{x-1})\sin(\theta_x) - \cos(\xi_x +\zeta_x)\cos(\theta_x)  \Big ]  = 0.~~~ 
\eea
Solving for $\omega$ we obtain,
\bea
\omega(k, \theta, \xi, \zeta) = \frac{1}{2}\Bigg [ {\mathfrak G}(\theta, \xi, \zeta, x)  \pm \sqrt{[{\mathfrak G}(\theta, \xi, \zeta, x)]^2 + 4\Big (k^2 {\mathfrak F}(\theta, \xi, \zeta, x) + k {\mathfrak Q}(\theta, \xi, \zeta, x) +{\mathfrak J}(\theta, \xi, \zeta, x)}\Big) \Bigg ].
\eea
Here
\bea
{\mathfrak G}(\theta, \xi, \zeta, x) &=&  \sin(\theta_x) \Big (  \sin( \xi_{x-1} - \zeta_{x-1} - \xi_{x}+\zeta_x ) \sin(\theta_{x-1}) \nonumber \\
&& + \sin(3\zeta_{x-1} +  \xi_{x} - \zeta_{x} + \xi_{x-1} )\cos(\theta_{x-1})\cot(\theta_{x-1}) \Big );\\
 {\mathfrak F}(\theta, \xi, \zeta, x) &=& \cos(\xi_x +\zeta_x)\cos(\theta_x) ; \\
  {\mathfrak Q}(\theta, \xi, \zeta, x) &=& -\sin ( 2\zeta_{x-1}+ \xi_{x} - \zeta_{x} ) \cot(\theta_{x-1})\sin(\theta_x) + \sin(\xi_x +\zeta_x)\cos(\theta_x) ; \\
{\mathfrak J}(\theta, \xi, \zeta, x) &=&   1+ \sin(\theta_x) \Big (  \cos(\xi_{x-1} - \zeta_{x-1} - \xi_{x}+\zeta_x   ) \sin(\theta_{x-1})  \nonumber \\
&&+\cos(3\zeta_{x-1} +  \xi_{x} - \zeta_{x} + \xi_{x-1}  )\cos(\theta_{x-1})\cot(\theta_{x-1}) \Big )  
\nonumber \\
&& -   \cos( 2\zeta_{x-1}+ \xi_{x} - \zeta_{x}  ) \cot(\theta_{x-1})\sin(\theta_x) - \cos(\xi_x +\zeta_x)\cos(\theta_x) ~~~
\eea
and the group velocity,
\bea
v_g^{SD}(\theta, \xi, \zeta) = \pm \frac{k \cos(\xi_x +\zeta_x)\cos(\theta_x) +  {\mathfrak Q}(\theta, \xi, \zeta, x)}{ \sqrt{[{\mathfrak G}(\theta, \xi, \zeta, x)]^2 + 4\Big (k^2 {\mathfrak F}(\theta, \xi, \zeta, x) + k {\mathfrak Q}(\theta, \xi, \zeta, x) +{\mathfrak J}(\theta, \xi, \zeta, x)}\Big)}.
\eea
By replacing $\theta_{x \pm 1}$, $\xi_{x\pm1}$ and $\zeta_{x\pm1}$   by $\theta$,  $\xi$, and $\zeta$,   we get back  the differential form and group velocity we obtained for  for the standard DTQW using SU(2) coin operation (section\,\ref{su2gv}).

\section{Iterative form for 2D DTQW using two-state particle}
\label{AppendixC}

The iterative form of the state of the particle at each position $(x, y)$ and time $t+1$ in the form of $\sigma_z$ Pauli basis states, $\psi^{\uparrow}$ and $\psi^{\downarrow}$ components at time $t$ are, 
\bea
\label{2ditA}
\psi^{\uparrow}(x, y, t+1) 
= \cos(\theta) \Bigg [ \cos(\vartheta) \Big  ( \psi^{\uparrow}_{x+1, y+1, t} + \psi^{\downarrow}_{x-1, y+1, t}  - \psi^{\downarrow}_{x-1, y-1, t} + \psi^{\uparrow}_{x+1, y-1, t} \Big )   \nonumber  \\
+ \sin(\vartheta) \Big ( \psi^{\uparrow}_{x+1, y+1, t} - \psi^{\downarrow}_{x-1, y+1, t}  - \psi^{\downarrow}_{x-1, y-1, t} - \psi^{\uparrow}_{x+1, y-1, t} \Big )  \Bigg ] \nonumber  \\
+ \sin(\theta) \Bigg [ \cos(\vartheta) \Big ( \psi^{\uparrow}_{x+1, y+1, t} + \psi^{\downarrow}_{x-1, y+1, t}  + \psi^{\downarrow}_{x-1, y-1, t} - \psi^{\uparrow}_{x+1, y-1, t} \Big ) \nonumber  \\
+ \sin(\vartheta) \Big ( \psi^{\uparrow}_{x+1, y+1, t} - \psi^{\downarrow}_{x-1, y+1, t}  + \psi^{\downarrow}_{x-1, y-1, t} + \psi^{\uparrow}_{x+1, y-1, t} \Big )  \Bigg ],  \\
\label{2ditB}
\psi^{\downarrow}(x, y, t+1) 
= -\sin(\theta) \Bigg [ \cos(\vartheta) \Big  ( \psi^{\uparrow}_{x+1, y+1, t} + \psi^{\downarrow}_{x-1, y+1, t}  - \psi^{\downarrow}_{x-1, y-1, t} + \psi^{\uparrow}_{x+1, y-1, t} \Big )   \nonumber  \\
+ \sin(\vartheta) \Big ( \psi^{\uparrow}_{x+1, y+1, t} - \psi^{\downarrow}_{x-1, y+1, t}  - \psi^{\downarrow}_{x-1, y-1, t} - \psi^{\uparrow}_{x+1, y-1, t} \Big )  \Bigg ] \nonumber  \\
+ \cos(\theta) \Bigg [ \cos(\vartheta) \Big ( \psi^{\uparrow}_{x+1, y+1, t} + \psi^{\downarrow}_{x-1, y+1, t}  + \psi^{\downarrow}_{x-1, y-1, t} - \psi^{\uparrow}_{x+1, y-1, t} \Big ) \nonumber  \\
+ \sin(\vartheta) \Big ( \psi^{\uparrow}_{x+1, y+1, t} - \psi^{\downarrow}_{x-1, y+1, t}  + \psi^{\downarrow}_{x-1, y-1, t} + \psi^{\uparrow}_{x+1, y-1, t} \Big )  \Bigg ].
\eea
\bea
\label{2ditA}
\psi^{\uparrow}(x, y, t+1) 
= \Big( \cos(\theta - \vartheta) +  \sin(\theta + \vartheta) \Big ) \psi^{\uparrow}_{x+1, y+1, t} 
+\Big( \cos(\theta + \vartheta) +  \sin(\theta - \vartheta) \Big ) \psi^{\downarrow}_{x-1, y+1, t} \nonumber \\
- \Big( \cos(\theta - \vartheta) -  \sin(\theta + \vartheta) \Big ) \psi^{\downarrow}_{x-1, y-1, t}
+ \Big( \cos(\theta - \vartheta) -  \sin(\theta + \vartheta) \Big ) \psi^{\uparrow}_{x+1, y-1, t},  \\
\label{2ditB}
\psi^{\downarrow}(x, y, t+1) 
= \Big( \cos(\theta + \vartheta) -  \sin(\theta - \vartheta) \Big ) \psi^{\uparrow}_{x+1, y+1, t} 
+\Big( \cos(\theta + \vartheta) -  \sin(\theta + \vartheta) \Big ) \psi^{\downarrow}_{x-1, y+1, t} \nonumber \\
+ \Big( \cos(\theta - \vartheta) +  \sin(\theta + \vartheta) \Big ) \psi^{\downarrow}_{x-1, y-1, t}
- \Big( \cos(\theta - \vartheta) +  \sin(\theta - \vartheta) \Big ) \psi^{\uparrow}_{x+1, y-1, t}.
\eea
The above expression cannot be decoupled like the way  the expressions were decoupled for 1D DTQW  to arrive at the continuum limit and the group velocity expression using dispersion relation. This restricts us to analytical study the localization of DTQW in 2D in the same lines as we did for the 1D DTQW.  However, this could lead to a completely new approach to arrive at the group velocity for 2D DTQW using two-state particle.
\vskip 0.3in
\end{widetext}

\end{document}